\documentclass[aps,apl,twocolumn,groupedaddress]{revtex4}
\input epsf
\usepackage{amsmath,amssymb}

\begin{document}


\title{Temperature and phase dynamics in superconducting weak-link}

\author{Anjan K. Gupta, Nikhil Kumar and Sourav Biswas}
\affiliation{Department of Physics, Indian Institute of Technology Kanpur, Kanpur 208016, India}

\date{\today}

\begin{abstract}
A time dependent thermal model for a superconducting constriction based weak-link (WL) is discussed for investigating the deterministic dynamics of its temperature and phase. A new dynamic regime is found where a non-zero voltage exists across the WL and its temperature stabilizes between the bath temperature and superconductor's critical temperature. This regime exists over a limited bias current range and gives rise to a new hysteretic regime in current-voltage characteristics. We also discuss the effect of fluctuations on the current-voltage characteristics and experimental implications of this dynamic regime.
\end{abstract}


\maketitle
\section{Introduction}
Superconducting weak-links (WLs) acting like Josephson junctions (JJs) \cite{likharev} have been used in micron-size superconducting quantum interference devices ($\mu$-SQUIDs), which in turn have been used in probing magnetism at small scales \cite{wernsdorfer,hasselbach-sc-squid}. Thus the physics of such WLs has been of interest and efforts have been made to understand their current-voltage characteristics (IVCs). The JJ like current-phase relation of the WL, which dictates the critical current variation with external flux in SQUIDs, is well understood using Ginzburg-Landau approach \cite{likharev}. However, the understanding of other detailed features in IVCs, particularly, hysteresis, its temperature evolution, and ways to eliminate it, in order to improve $\mu$-SQUID's performance, needs further work.

Hysteresis also exists in the I-V characteristics of the conventional JJs due to large junction capacitance ($C$). This is modeled by resistively and capacitively shunted junction (RCSJ) model \cite{McCumber,tinkham-book} and thus the hysteresis in these JJs can be eliminated by using a small-value shunt resistor ($R$) in parallel with the JJ. In the non-hysteretic regime of these JJs, for bias-currents higher than I$_c$ the current is dynamically shared between the shunt resistor and the JJ. In WLs, the hysteresis is observed at low temperatures while at higher temperatures it is found to disappear \cite{dibyendu-prb,sophie-prb,sayanti,hazra-thesis}. Two approaches regarding hysteresis in these WLs have been proposed: 1) hot-spot model \cite{scockpol} and 2) RCSJ like approach but with an effective time constant, similar to capacitor charging time $RC$ of RCSJ model but, related to the recovery of the superconducting order parameter in the WL \cite{song}. Direct evidence of hot-spot in some WLs has been found in hysteretic regime \cite{herve-prl-hot-spot}. The second approach explains the I-V curves in the hysteretic regime by asserting that the superconductivity recovery (or Cooper-pair relaxation) time is larger than typical phase evolution time. Kramer and Watts-Tobin \cite{kramer}, using time dependent Ginzburg-Landau (TDGL) equations, found hysteretic IVCs in superconducting filaments due to oscillatory phase-slip solutions. The hysteresis dominates in certain parameter regime \cite{vodolazov-IVC} where relaxation time of the magnitude of the order-parameter, i.e. $\tau_{|\psi|}$, is greater than that of the gradient of its phase, i.e. $\tau_{\phi}$. The non-hysteretic I-V curves, observed at higher temperatures, using this approach have often been modeled using an over-damped RCSJ model \cite{sophie-prb}.

The hot-spot model was proposed as a static thermal model \cite{scockpol}, where temperature profile near the WL is assumed to be time-independent for a given bias current. While the features in hysteretic regime are satisfactorily understood with the static thermal model it seems reasonable that a non-hysteretic IVC, seen at higher temperatures \cite{dibyendu-prb,sophie-prb,sayanti,hazra-thesis}, will have a current switching between the resistive and non-resistive branches in a dynamic fashion as is the case with the RCSJ  model or with phase-slip processes \cite{kramer}. A phase-slip can amount to cooling \cite{schmid-qp-cooling} or heating \cite{vodolazov-qp-heating} of quasi-particles in the WL, affecting its current-voltage characteristics, but it will always lead to a net heating in the region near WL over the branch-imbalance length \cite{branch-imb}. The thermal dynamics due to phase-slips and its effect on the order-parameter evolution has not been explicitly investigated by such time-dependent models. In one model by Vodolazov et. al. \cite{vodolazov-radiation} heat evacuation was analyzed but the effect of thermal relaxation time-scale on the evolution of the order parameter magnitude was not described. Shah et. al. \cite{goldbart} proposed a time-dependent thermal model for the stochastic phase-slips and evacuation of heat thus generated in long superconducting nano-wires. The dynamics of local temperature was thus analyzed and it was found that in one regime phase-slips can cause a thermal runaway while in another regime one can have occasional phase-slips with some local heating but without a runaway. Thermal stability of local superconducting order in presence of super-current is of great practical importance, in particular for SC-magnet wires and cables, helium level sensors, radiation-detectors \cite{berdiyo-apl}, SC-WLs and other small scale SC structures \cite{shah-nano-wire,berdiyo-prl}.

In this paper we discuss a similar time-dependent thermal model to investigate the deterministic dynamics of phase and temperature in a superconducting WL. We find a new dynamic regime, as a result of the heat balance between heat evacuation and generation. In this regime, superconducting phase-difference across the WL evolves continuously giving rise to a DC voltage and raising the WL temperature but without a thermal runaway. A fraction of bias-current, on average, is still carried as the super-current in this regime. This regime exists over a limited bias-current range and gives rise to a new thermal hysteretic regime. Effect of fluctuations and experimental consequences are also discussed briefly.
\section{time dependent thermal Model}
\begin{figure}
\epsfxsize = 3 in \epsfbox{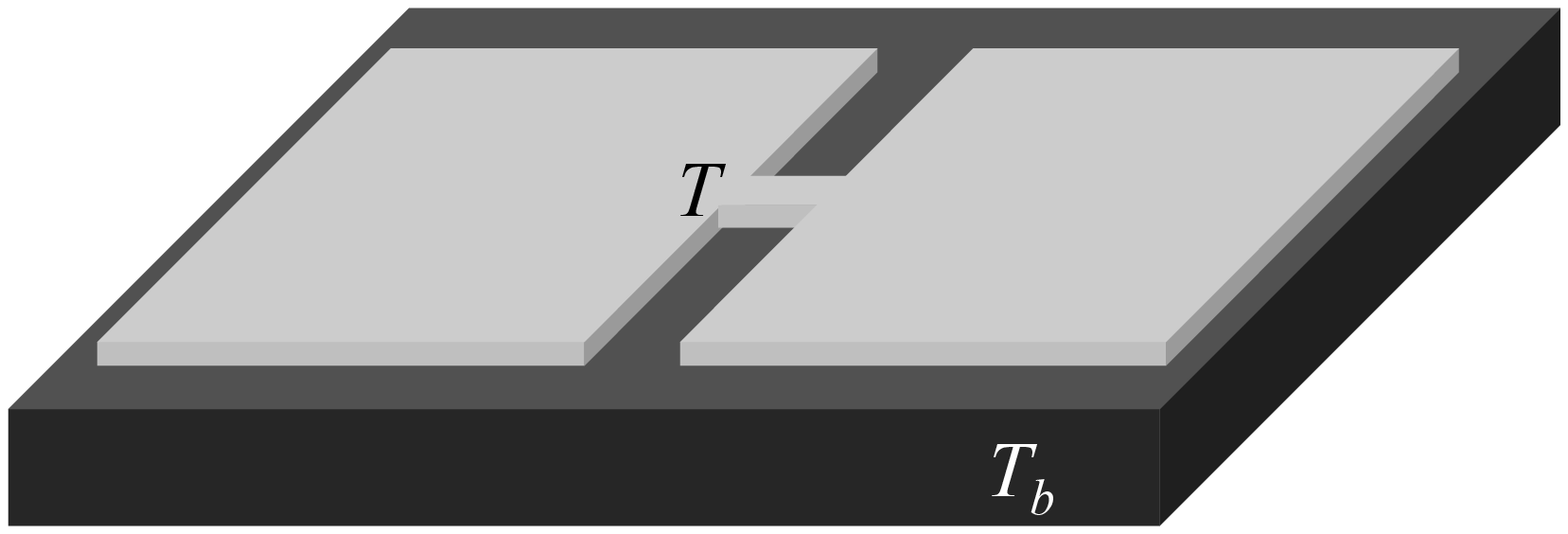}
\caption{\label{fig:schematic}Schematic of a superconducting constriction based WL on a substrate held at bath temperature T$_b$. The constriction's instantaneous temperature (T) can be different from T$_b$ depending on the bias current I$_b$ as discussed in the text.}
\end{figure}

We first recall the static thermal model as proposed by Scockpol et. al. \cite{scockpol}. In this model when the current is ramped down, across a superconducting constriction (see fig. \ref{fig:schematic}), from a value above its critical current ($I_c$), a static hot-spot, with temperature greater than $T_c$, is sustained near the constriction down to a current, which we call as static re-trapping current, $I_{sr}$. This defines a bistable region (for $I_c>I_{sr}$) in the I-V characteristics in bias current range $I_{sr}<I<I_c$ with one branch having zero voltage and the other a finite voltage. The finite voltage branch is found by solving the static heat conduction equation so as the heat generated by the resistive hot spot, at temperature greater than $T_c$, is conducted away to the substrate or the bulk superconductor nearby. For constriction-dimensions less than thermal healing length, $\eta=\sqrt{\kappa t/\alpha}$, whole of the constriction stays above $T_c$, in resistive-state, and the $T=T_c$ interface occurs in the bulk electrodes. Here, $\kappa$ is thermal conductivity of the film, $t$ is film thickness and $\alpha$ is the heat-transfer coefficient between film and the substrate kept at bath temperature $T_b$. As the bath temperature rises, both $I_c$ and $I_{sr}$ decrease but with a different temperature dependence and thus the two cross each other at some value of the bath temperature, $T=T_h$ \cite{dibyendu-prb} giving rise to a non-hysteretic behavior above $T_h$.

We consider the same superconducting-constriction as a WL, between two bulk superconductors (see fig.\ref{fig:schematic}), with its dimensions smaller than $\eta$ \cite{scockpol} so that the whole WL can be assumed to be at the same temperature $T$. When a bias current ($I$) flows through the WL, its instantaneous temperature ($T$) can be larger than the bath temperature ($T_b$) giving rise to the heat flow from the WL to the bath via the interface with the substrate or via the bulk superconducting electrodes on the two sides. The temperature will relax over the length scale $\eta$ in the two superconducting electrodes.

Another consequence of this heating is the reduction in the critical current, $I_c(T)$, of the WL. The bias current, $I$, will be dynamically shared between normal resistance, $R_N$ and the super-current ($I_s$), thus $I=(V/R_N)+I_s$. We also assume $R_N$ to be independent of temperature. The resistivity at such low temperatures is independent of temperature, so the main assumption here is the confinement of the hot-spot to the constriction. The dynamic current sharing, here, is quite similar to the RCSJ model with the additional fact that the heat generated in resistive shunt heats up the WL. A time dependent voltage, $V(t)$, arises across the WL whenever $\phi$, i.e. the phase difference between the two superconductors across the WL, changes with time. This is described by the ac-Josephson relation, $V=(\hbar/2e)d\phi/dt=(\Phi_0/2\pi)d\phi/dt$ with $\Phi_0=h/2e$ as the flux quantum. This voltage $V$ drives a normal current through the WL. We assume a (quasi-static) diffusive limit so that the mean free time for electron scattering is smaller than the time scale over which $V$ or $\phi$ changes. We further take a short WL limit so that super-current ($I_s$) and $\phi$ relation is sinusoidal, i.e. $I_s=I_c(T)\sin\phi$ \cite{likharev}. Thus at a given instant of time, the total bias current, shared between resistive current and super-current branch is given by,
\begin{eqnarray}
I=\frac{V}{R_N}+I_c(T)\sin\phi
\label{eq-current-bal}
\end{eqnarray}
We further assume that the rate of heat evacuation from the WL is linear with $(T-T_b)$. With $k$ as an effective heat loss coefficient independent of temperature, we write the heat evacuation rate as $k(T-T_b)$. The heat loss from the WL occurs through the bulk electrodes and also via the interfaces. As we saw earlier, a time evolving $\phi$ gives rise to $V$, which enforces a resistive current giving rise to heat generation in the WL at a rate $V^2/R_N=(\Phi^2_0/4\pi^2 R_N)(d\phi/dt)^2$. The rate at which the WL temperature changes with time is governed by the net rate of heat accumulation in the WL and the heat capacity of the WL, i.e. $C_{WL}$. We assume $C_{WL}$ as temperature independent. Thus, the expression governing the evolution of the WL temperature is given by,
\begin{eqnarray}
C_{WL}\frac{dT}{dt}&=&-k(T-T_b)+\frac{V^2}{R_N}\nonumber \\
&=&-k(T-T_b)+\frac{\Phi^2_0}{4\pi^2 R_N}\left(\frac{d\phi}{dt}\right)^2
\label{eq-heat-bal}\end{eqnarray}
We note that the branch-imbalance and related length scale will also affect $C_{WL}$. When the current is carried by quasi-particles across the WL their relaxation to Cooper-pairs and subsequent heat generation will occur \cite{branch-imb} over branch-imbalance length going beyond the WL. This will effectively increase $C_{WL}$.

The heat generation can also be looked at as the release of energy with each phase-slip event, i.e. a phase change of $2\pi$. These phase-slips may occur at certain rate. The energy released in each phase-slip event, at a given bias-current, $I$, is given by $(\Phi_0/2\pi)\int_{0}^{2\pi} I d\phi$ i.e. $I\Phi_0$. The time-scale, over which this phase-slip occurs, is of order $\tau_J=\frac{\Phi_0}{R_N I_c(T_b)}$, which we shall call as Josephson time. Thus the energy released in each phase-slip is independent of $R_N$ but the rate of phase-slip depends on $R_N$. Here we have also assumed that the charge-imbalance time, associated with quasi-particle to condensate conversion or vice-versa, and the Ginzburg-Landau time, which describes the relaxation of superconducting order-parameter, are smaller than $\tau_J$.

We further simplify above equations by defining a dimensionless normalized temperature $p=(T-T_b)/(T_c-T_b)$, thermal time constant $\tau_{th}=C_{WL}/k$ and normalized bias current $i_b=I/I_{c}(T_b)$. We also non-dimensionalize time using $\tau_J$ by defining $\tau=t/\tau_J$. Thus from eqs. \ref{eq-current-bal} and \ref{eq-heat-bal} we get,
\begin{eqnarray}
\dot{\phi}=2\pi[i_b-i_c(p)\sin\phi]
\label{eq-phase}
\end{eqnarray} and
\begin{eqnarray}
\alpha\dot{p}=-p+\frac{\beta}{4\pi^2}\dot{\phi}^2
\label{eq-temperature}
\end{eqnarray}
Here $i_c(p)=I_c(T)/I_c(T_b)$, $\alpha=\tau_{th}/\tau_J$ and $\beta=\frac{I^2_{c}(T_b)R_N}{k (T_c-T_b)}$. For most experimentally studied devices it turns out that $\alpha>>1$, i.e. $\tau_{th}>>\tau_J$. The bath temperature dependent parameter $\beta$ is an important parameter as it controls the phase and temperature dynamics. We can interpret $\beta$ by writing it as a ratio of two time scales in two different ways: 1) $\beta=\tau_{th}/\left(\frac{C_{WL}(T_c-T_b)}{I^2_{c}(T_b)R_{N}}\right)$ or 2) $\beta=\left(\frac{2 \pi E_J}{k (T_c-T_b)}\right) /\tau_J$ with $2\pi E_J=\Phi_0 I_c(T_b)$. In the first case it represents the ratio of the heat conduction time ($\tau_{th}$) to the time required for heating the WL by $\Delta T=T_c-T_b$ using the resistive Joule power at critical current. The second expression represents the ratio of the time taken to conduct away the heat generated (at $I=I_c$) in one phase-slip event to the phase-slip time. In either way $\beta$ is a measure of the competition between heat generation and heat evacuation with $\beta>1$ representing the dominance of heat generation. It is somewhat like the parameter $\gamma$ used in TDGL equation approach \cite{kramer,vodolazov-IVC}, which depends on $\tau_{|\psi|}$ and $\tau_{\phi}$.

In order to find the DC $I-V$ characteristics of a WL, we need to find the steady state solutions of non-linear eqs. \ref{eq-phase} and \ref{eq-temperature} and their stability with respect to fluctuations. Two types of steady states are possible: 1) static state where temperature and phase take time-independent values, 2) oscillatory or limit cycle type steady state where $\dot{\phi}$ and temperature oscillate about an average value. We have been guided by numerical solutions, illustrated later for a special case, of the above non-linear equations which eventually led us to analytical solutions for large $\alpha$ values which is true for most studied devices.

\section{Steady-State Solutions}
For static solutions, $\dot{p}=0$ and if $p<1$, $\dot{\phi}=0$. We have two such solutions. In first case we get $i_b=i_c(p)\sin\phi$ and $p=p_s=0$, i.e. $I=I_c(T_b)\sin\phi$ and $T=T_b$. This solution is possible only for $I<I_c(T_b)$ (or $i_b<i_c(0)$). The second static solution is possible only for $T>T_c$, i.e. $p>1$, and thus $i_c(p)=0$ and so no super-current flows through the WL. In this case $\phi$ is irrelevant and $d\phi/dt$ (=$2\pi V/\Phi_0$) can be thought in terms of the voltage across the junction. By eliminating $\dot{\phi}$ from eq. \ref{eq-phase} and \ref{eq-temperature}, we get for static temperature $p_s=\beta i^2_b$, i.e. $T_s=T_b+(I^2 R_N/k)$. The second solution exists only for currents for which $T>T_c$ (or $p>1$), i.e. $I>I_{sr}$ (or $i_b>i_{sr}$) with
\begin{eqnarray}
I_{sr}=\sqrt\frac{k(T_c-T_b)}{R_N} \text{ or } i_{sr}=\frac{1}{\sqrt{\beta}}
\label{eq-Isr}
\end{eqnarray} We note that the static re-trapping current, $I_{sr}$, is bath temperature dependent but it is independent of critical current. From here we see that when $\beta>1$, for $I_{sr}(T_b)<I<I_c(T_b)$ at a given $T_b$ we have two possible static solutions. There exists a cross-over temperature, $T_b=T_h$, at which $I_c$ and $I_{sr}$ are equal (or $i_{sr}=1$), given by $\beta=1$. For $T_b<T_h$, $I_c>I_{sr}$ so we would get hysteretic IVCs while for $T_b>T_h$, $I_c<I_{sr}$ we get non-hysteretic IVCs. This $T_h$ is determined by the temperature dependence of $I_c$. This $T_h$ also brings out the significance of the parameter $\beta$ with $\beta=1$ (at $T_h$) as the cross-over point between the relative magnitudes of heat generation and heat evacuation.

\begin{figure}
\epsfxsize = 2.5 in \epsfbox{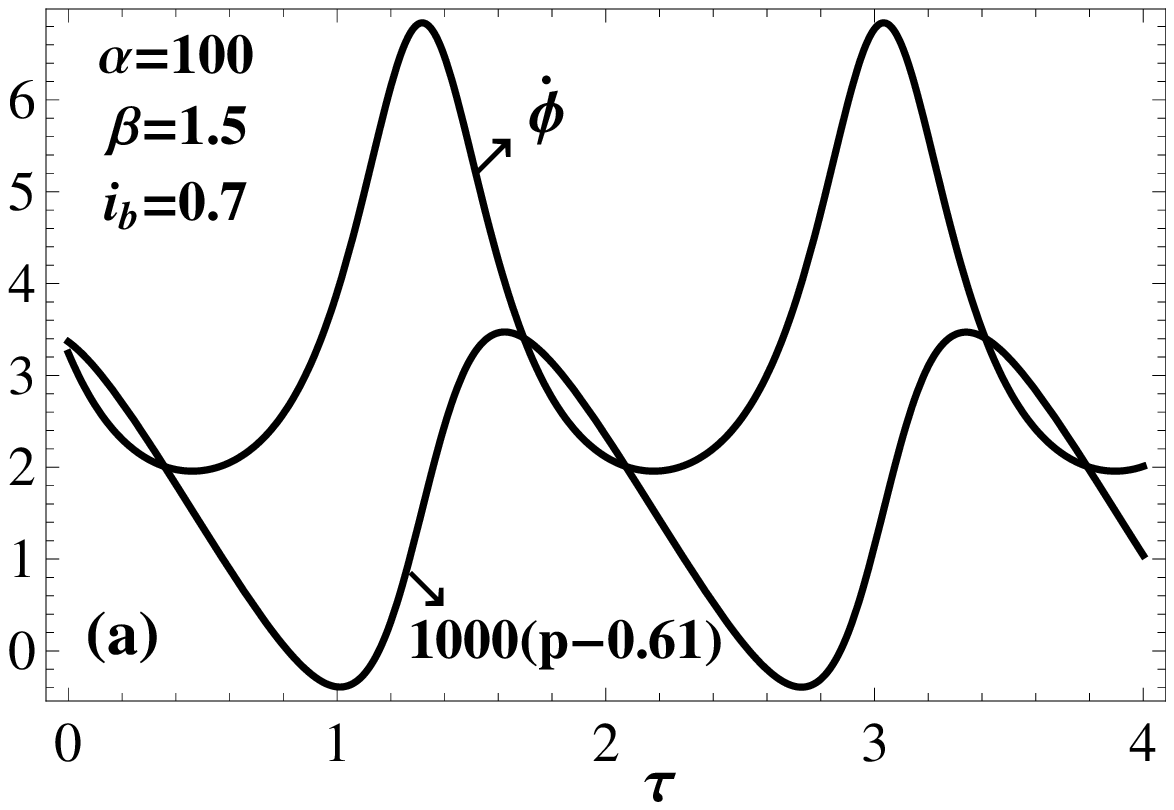}
\epsfxsize = 2.4 in \epsfbox{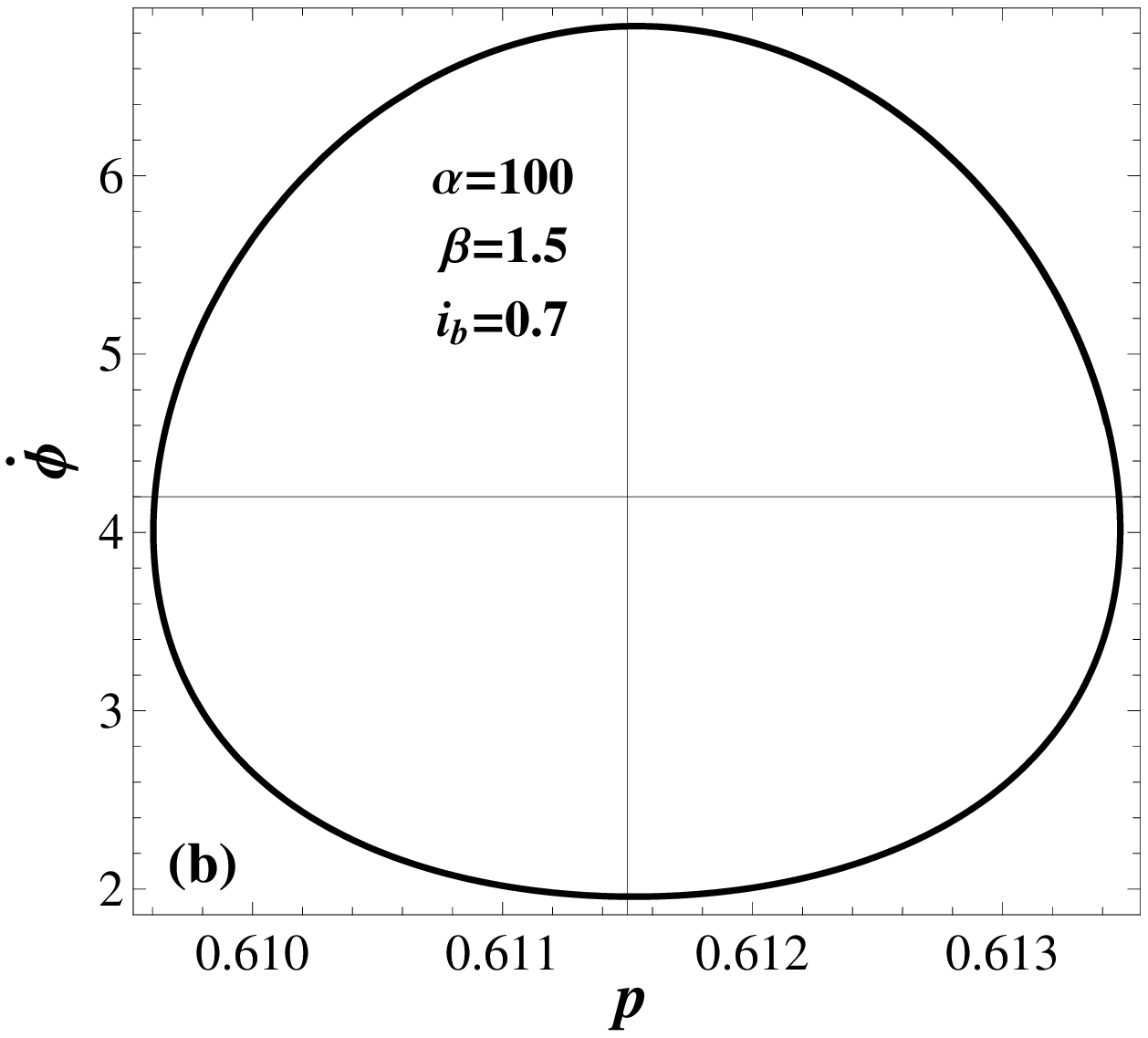}
\caption{\label{fig:num-soln} {\bf (a)} Oscillatory steady state time evolution of $p$, actually $1000(p-0.61)$, and $\dot{\phi}$ for $\alpha=100$, $\beta=1.5$ and $i_b=0.7$ found by numerically solving eq. \ref{eq-phase} and \ref{eq-temperature}. {\bf (b)} shows the steady-state trajectory (limit-cycle) in $p-\dot{\phi}$ space.}
\end{figure}
Next we look for oscillatory steady state (or stable limit-cycle solutions) in the limit of $\alpha>>1$. In this limit we find that the heat generated by a phase-slip event, i.e. $I\Phi_0$, over time $\tau_J$ gets evacuated over a much larger time scale $\tau_{th}$. Thus we make an approximation, which we shall justify later, that the change in temperature, and thus change in $i_c$, during a phase-slip is negligible. In this limit, we can take $p$ to be fixed while integrating eq. \ref{eq-phase} over one phase-slip happening over time $\tau_{ps}$ (in units of $\tau_J$). Using eq. \ref{eq-phase}, $\tau_{ps}$  is given by
\begin{eqnarray}
\tau_{ps}=\int_0^{2\pi}\frac{d\phi}{2\pi[i_b-i_c(p)\sin\phi]}=\frac{1}{\sqrt{i^2_b-i^2_c(p)}}.
\label{eq-tau-ph}
\end{eqnarray}
Using eq. \ref{eq-phase}, we also get, $\langle\dot{\phi}^2\rangle_{\tau_{ps}}=\frac{1}{\tau_{ps}} \int_0^{\tau_{ps}}\dot{\phi}^2 d\tau =\frac{1}{\tau_{ps}}\int_0^{2\pi}\dot{\phi}d\phi =\frac{2\pi}{\tau_{ps}}\int_0^{2\pi}\left[i_b-i_c(p)\sin\phi\right] d\phi =\frac{4\pi^2 i_b}{\tau_{ps}}$. We use this in eq. \ref{eq-temperature} to get, $\alpha\langle\dot{p}\rangle_{\tau_{ps}}=\alpha\Delta p/\tau_{ps}=-p+\beta i_b\sqrt{i^2_b-i^2_c(p)}$. Combining with the previously discussed static solutions, we write from eq. \ref{eq-temperature}
\begin{widetext}\begin{eqnarray}
\alpha\langle\dot{p}\rangle_{\tau_{ps}}=-\frac{dU}{dp}=\left\{\begin{array}{lll}
&-p &\mbox{for }i_b\leq i_c(p)\\
&-p+\beta i_b\sqrt{i^2_b-i^2_c(p)}& \mbox{for }i_b>i_c(p) \mbox{ \& } p\leq1 \\
&-p+\beta i_b^2 & \mbox{for }p>1\end{array}\right.
\label{eq-avg-dp-dtau}
\end{eqnarray}\end{widetext}
Here $U(p)$ defines a fictitious potential describing the dynamics of $p$ for whole range of $p$-values. For steady-state behavior of $p$ we have to analyze the extremum points of $U(p)$. The minima of $U(p)$ describe the stable steady-states while a maximum separating the two minima gives the sensitivity of the stable steady-state to fluctuations. One can readily see from eq. \ref{eq-avg-dp-dtau} that these extremum points are given by $p=0$, $\beta i_b\sqrt{i^2_b-i^2_c(p_0)}$ and $\beta i_b^2$.

To find the magnitude of temperature oscillations, we recall that the heat evacuated during a single phase-slip event is very small as compared to the heat generated. If we neglect the heat evacuated during one phase-slip event we can estimate the temperature rise as $\Delta T=I\Phi_0/C_{WL}=i_b \frac{E_J}{C_{WL}}=i_b\frac{\beta}{\alpha}(T_c-T_b)$ or $\Delta p=i_b \beta/\alpha$. This is actually an upper limit as some heat evacuation will happen during phase-slip. So for small values of $\beta/\alpha$ the oscillation in temperature will be small. We note that this argument is valid if change in critical current is small due to this small change in temperature. In case the critical current declines very fast with temperature this approximation will become invalid.

\begin{figure}
\epsfxsize = 2.8 in \epsfbox{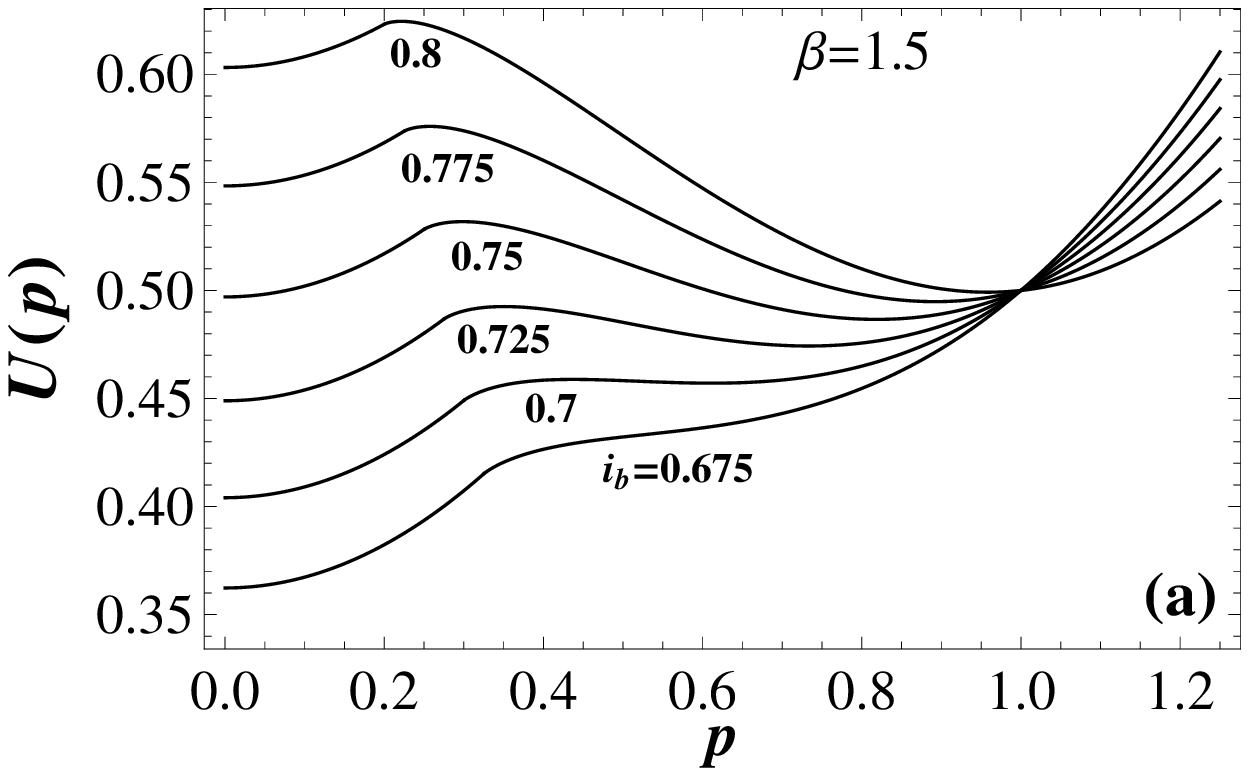}
\epsfxsize = 2.8 in \epsfbox{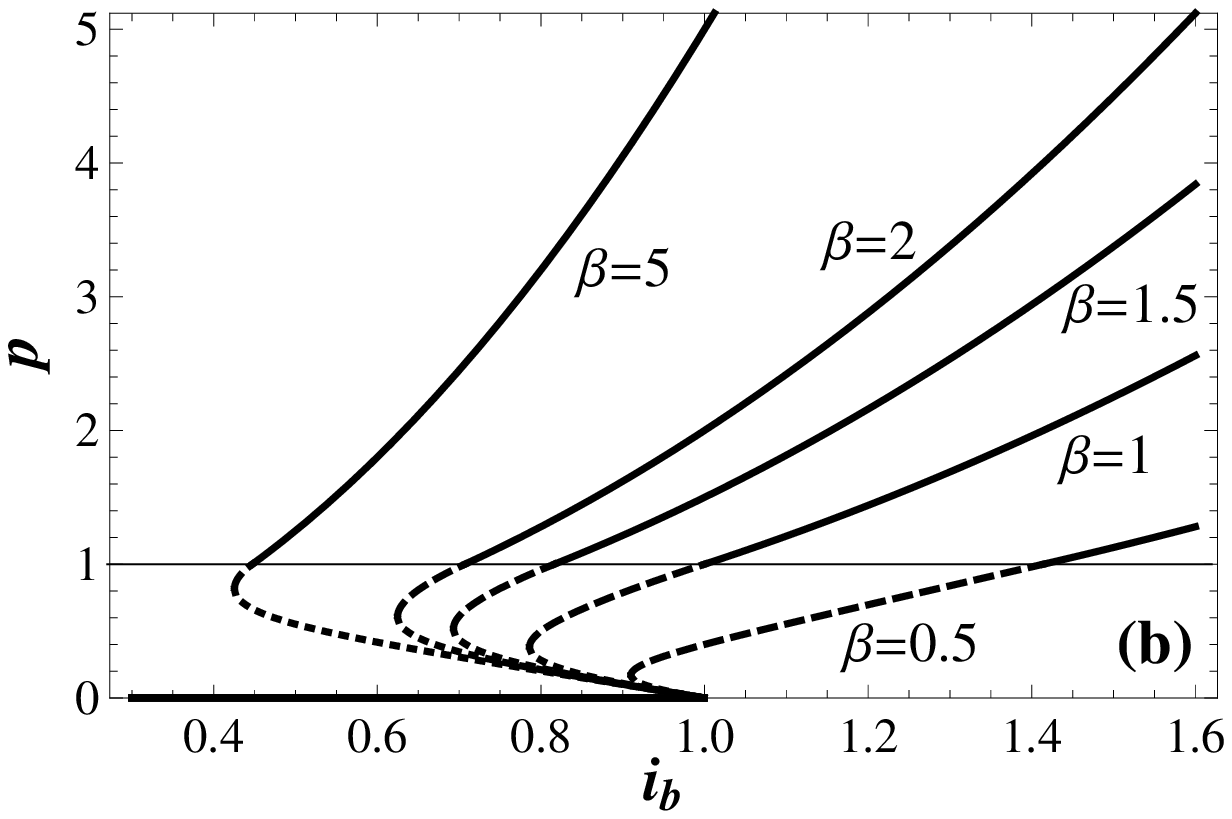}
\caption{\label{fig:p-solns} {\bf (a)} The fictitious potential $U(p)$ for $\beta=1.5$ for various values of $i_b$. Here we see that for $i_b<i_{dr}=0.693$ [see eq.\ref{eq:Idr}] there is only one minima at $p=0$ and for $1>i_b>i_dr$ there are two minima. For $i_b>i_{sr}=0.816$ [see eq.\ref{eq-Isr}] the higher $p$ minima occurs at $p>1$ i.e. the WL is fully normal. {\bf (b)} Shows the evolution of these minima, $p_s$ (continuous lines), $p_+$ (dashed lines) and maxima, i.e. $p_-$ (dotted lines), with $i_b$ for various $\beta$ values.}
\end{figure}
After knowing the steady-state temperature $p_0$ one can find the IVCs. For $p_0<1$, the DC voltage is given by $V=\langle V(t)\rangle_{\tau_{ps}}=\frac{\Phi_0}{2\pi\tau_J}\langle\dot{\phi}\rangle_{\tau_{ps}}=
\frac{\Phi_0}{2\pi\tau_J}\frac{2\pi}{\tau_{ps}}=R_N I_c(T_b)\sqrt{i^2_b-i^2_c(p_0)}$. For $p_0\geq 1$, $V=R_N I$. Combining, we get
\begin{eqnarray}
V=\left\{\begin{array}{lll}
&R_N\sqrt{I^2-I^2_c(T_0)}& \mbox{for }T_0<T_c \\
&R_N I & \mbox{for }T_0 \geq T_c\end{array}\right..
\label{eq-IV}
\end{eqnarray} This expression resembles that of the overdamped RCSJ except for the value of $T_0$, which is dependent on $I$ and $T_b$.
\section{Results for linear $I_c(T)$}

\begin{figure}
\epsfxsize = 3 in \epsfbox{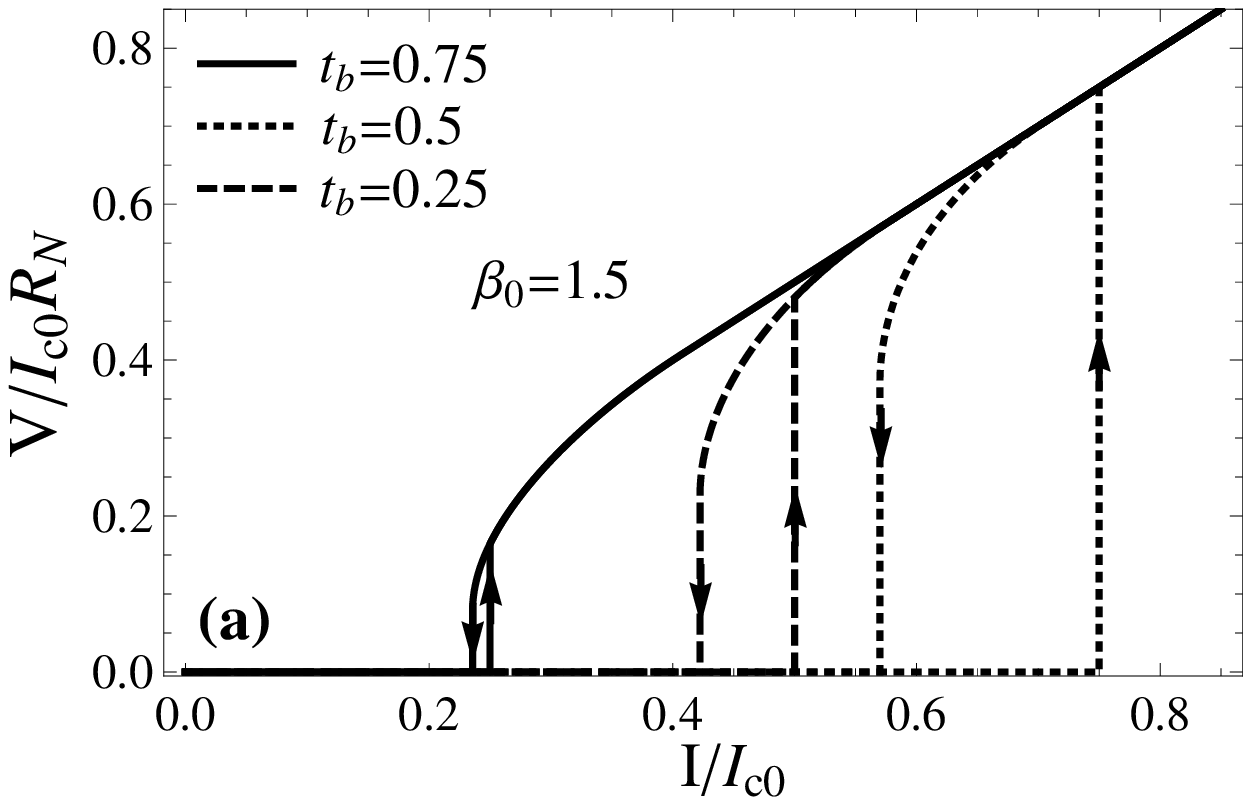}
\epsfxsize = 3 in \epsfbox{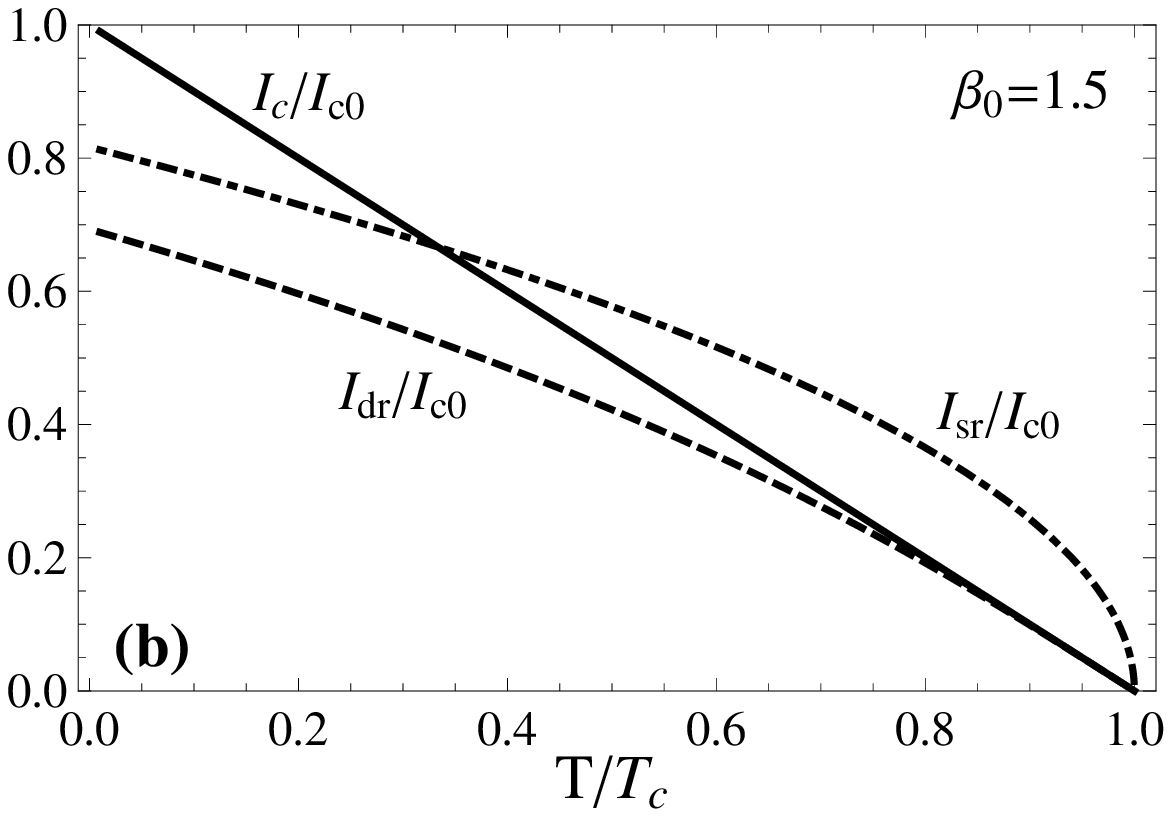}
\caption{\label{fig:IVC}{\bf (a)} IVCs for $\beta_0=1.5$ at three different $t_b$ values. {\bf (b)} Expected variation of $I_{dr}$, $I_{sr}$ for linear $I_c(T)$ (also shown).}
\end{figure}
Here we illustrate the above phase and temperature dynamics for short superconducting WLs where $I_c(T)$ varies linearly with temperature \cite{likharev}. Thus we assume $I_c(T)=I_{c0}[1-(T/T_c)]\Theta[1-(T/T_c)]$ with $\Theta$ as the step-function. Thus we get $i_c(p)=(1-p)\Theta(1-p)$ and $\beta=\beta_0\left(1-\frac{T_b}{T_c}\right)$ with $\beta_0=\frac{I^2_{c0}R_N}{k T_c}$ and $T_h=T_c\left(1-\frac{1}{\beta_0}\right)$. With this we can solve eq. \ref{eq-phase} and \ref{eq-temperature} numerically with given initial conditions. Fig. \ref{fig:num-soln} shows the oscillatory steady state time evolution of $p$ and $\dot{\phi}$ for $\alpha=100$, $\beta=1.5$ and $i_b=0.7$. This regime has similarities to the one found by TDGL equations \cite{kramer,vodolazov-IVC} but in our approach the phase-slip oscillations are driven by the thermal dynamics. This steady state is seen only for certain initials conditions, as discussed later, and for others we get steady state as $p=0$ and $\dot{\phi}=0$. $p$-oscillates from its mean value of about 0.6115 by an amplitude less than 0.002 justifying our earlier claim of small temperature oscillations for large $\alpha$ values. One can also see that the oscillations are non-sinusoidal and $\dot{\phi}$ and $p$ oscillate in quadrature. This is expected as the rate of change of temperature is maximum at the peak of $\dot{\phi}$ where the heat (power) generated is highest while at the minimum of $\dot{\phi}$ the heat evacuation dominates (see eq. \ref{eq-temperature}).

In order to get more general idea about the dynamics of $p$ (and $\phi$) we use $i_c(p)$ expression in eq. \ref{eq-avg-dp-dtau} to get,
\begin{widetext}
\begin{eqnarray}
U(p)=\left\{\begin{array}{lll}
&\frac{p^2}{2}+\frac{\pi\beta i^3_b}{4} &\mbox{for }p\leq(1-i_b)\\
&\frac{p^2}{2}+\frac{\beta i_b}{2}\left[i^2_b\sin^{-1}\left(\frac{1-p}{i_b}\right)+(1-p)\sqrt{i^2_b-(1-p)^2}\right]& \mbox{for }(1-i_b)<p\leq1 \\
&\frac{p^2}{2}-\beta i_b^2(p-1) & \mbox{for }p>1\end{array}\right.
\end{eqnarray}
\end{widetext}
This $U(p)$ depends on $T_b$ through $\beta$. We plot $U(p)$ in fig. \ref{fig:p-solns}(a) for $\beta=1.5$ for selected values of $i_b$. $U(p)$ has at least one minimum at $p=0$ for $i_b<1$, which becomes a maxima if $i_b>1$. A second minima exists for $i_b>i_{dr}$ with the dynamic re-trapping current, $i_{dr}$, given by \begin{eqnarray}i^2_{dr}=(\sqrt{1+4\beta^2}-1)/2\beta^2.\label{eq:Idr}\end{eqnarray} For $1>i_b>i_{dr}$, $U'(p)$ admits two zeroes at $p_{\pm}=(i^2 \beta^2 \pm \sqrt{-i^2 \beta^2 + i^4 \beta^2 + i^6 \beta^4})/(1 + i^2 \beta^2)$ other than $p=0$. Here $p_+$ corresponds to the minimum describing a stable limit-cycle while $p_-$ is the maximum of $U(p)$. We plot, in fig. \ref{fig:p-solns}(b), the temperatures, $p_s$, $p_+$ and $p_-$, at which the minima and maxima occur as a function of $i_b$ for various $\beta$ values. We clearly see a bistable region for $1>i_b>i_{dr}$.

In order to find the IVCs for a given $\beta_0$ value we have three different segments that describe the steady-state solutions. These are, 1) $i_{dr}>i_b>0$: static superconducting, 2) $i_{sr}>i_b>i_{dr}$: dynamic steady state and 3) $i_b>i_{sr}$: static resistive. Fig. \ref{fig:IVC}(a) shows the expected IVCs for a constriction with $\beta_0=1.5$. We see a non-linear IVC just above $I_{dr}$ as a fraction of current is super-current and above $I_{sr}$ the IVCs become linear. Non-linearity in IVCs is also found in static thermal models due to spread of resistive hot-spot with increasing temperature, however, in that case no super-current flows above the re-trapping current. We have not captured this static non-linearity due to the simplification that only the constriction heats up. Fig. \ref{fig:IVC}(b) shows the temperature variation of the three currents, namely $I_c$, $I_{dr}$ and $I_{sr}$, for $\beta_0=1.5$. We see that $I_c$ and $I_{sr}$ cross each other at $T/T_c=1-1/\beta_0$. $I_{dr}$ always stays below $I_c$ and they both go to zero at $T_c$ but the difference between the two becomes increasingly small as they approach zero.

\section{Discussion and Conclusions}
$U(p)$ plot in fig. \ref{fig:p-solns}(a) also captures the sensitivity of the two stable points to fluctuations or initial conditions. Thus when the system is in one of the two minima and if a fluctuation causes it to overcome the maximum separating the two minima, it can transit to the other minimum. The minimum energy required to overcome the maximum so as to go from $p=0$ minimum to the other is $C_{WL}(T_c-T_b)p_-$ while for reverse process the system has to loose at least $C_{WL}(T_c-T_b)(p_+-p_-)$ energy. The barrier may be overcome due to stochastic phase-slip processes \cite{goldbart}. More work is needed to fully capture the role of stochastic fluctuation on superconducting constrictions. Similar to RCSJ model, the fluctuations will give rise to a distribution in the observed critical and re-trapping currents and in case of large fluctuations the bistable region may be completely wiped out. In fact as one approaches $T_c$ this barrier, between two stable states, becomes smaller while $I_c$ and $I_{dr}$ approach each other and thus the hysteresis will disappear, due to fluctuations, before $T_c$. The maxima (or $p_-$) also describes the sensitivity to the initial conditions for solving eq. \ref{eq-phase} and \ref{eq-temperature}. Thus if one starts with an initial $p$-value on the left of the maxima the system will stabilize in the minima on the left after a transient. This has been confirmed by numerical solutions.

We also see in fig. \ref{fig:p-solns}(b) that the dynamic region (i.e., $i_{dr}>i_b>i_{sr}$) shrinks as $\beta$ becomes larger. This will be the case for WLs with high $I_c$ values. For WLs with very small $I_c$, more precisely $I_{c0}<kT_c/R_N$ so that $\beta_0<1$, the static re-trapping current ($I_{sr}$) will be smaller than $I_c$ at all temperatures; however the hysteresis will persist as $I_{dr}<I_c$ [see fig. \ref{fig:IVC}(b)] at all temperatures. For very large $\beta$ (or $I_c$) the dynamic region may be completely suppressed and no super-current will exist in the finite voltage branch. Thus for SQUIDs with constrictions replacing Josephson Junctions one will see oscillations in the re-trapping current for small $\beta_0$ (or $I_c$) as is the case in some of the experiments \cite{sayanti,sophie-prb} and not others \cite{hazra-thesis}, where $I_{c}$, and therefore $\beta$, is large.

The thermal time constant, i.e. $\tau_{th}$, can also be interpreted as a time-scale over which the constriction cools below $T_c$ and so the superconducting order-parameter is restored over this $\tau_{th}$. This healing time for superconductivity will compete with the other superconductivity recovery time-scales such as $\tau_{|\psi|}$ discussed earlier.
This will be relevant in cases where the thermal healing time is small. If $\tau_{th}$ is smaller than other time-scales, like $\tau_{|\psi|}$, relevant for recovery of superconductivity in the WL, the hysteresis and detailed I-V characteristics will not be completely dictated by the temperature dynamics and TDGL approach \cite{kramer,vodolazov-IVC} may be more appropriate in such cases. If $\tau_{th}$ dominates the other time scales, the superconducting order parameter cannot recover faster than $\tau_{th}$ and thus TDGL alone cannot describe the order-parameter dynamics and I-V characteristics. In such a case the thermal dynamics alone may be sufficient to deduce the I-V characteristics of the WL.

In conclusion, we have illustrated the deterministic dynamics of temperature and phase in a superconducting constriction using a simple time dependent thermal-model. A new dynamic region is found where the bias-current is shared dynamically between super-current and normal current giving rise to a finite voltage. The Joule heat thus generated in the constriction region raises its temperature but not above $T_c$ in this regime. A new bi-stable region is found between $I_{dr}$ and $I_{sr}$. The oscillations of re-trapping current in some of the WL based SQUIDs seen in earlier reported experiments \cite{sayanti,sophie-prb} is consistent with this dynamic regime.

\section{Acknowledgements}
AKG would like to acknowledge a research grant from the CSIR of the Govt. of India. We thank Herv\'{e} Courtois for his comments on the manuscript. We would also like to thank the referee for bringing the references \cite{kramer,vodolazov-IVC,vodolazov-qp-heating,vodolazov-radiation,berdiyo-apl,berdiyo-prl}, to our notice.

\end{document}